\documentclass[twocolumn,showpacs,preprintnumbers]{revtex4}
\usepackage{amssymb}
\usepackage{amsfonts}
\usepackage{float}
\usepackage{graphicx}
\usepackage{dcolumn}
\usepackage{bm}
\usepackage{epsfig}
\usepackage{amsmath}
\usepackage{subfigure}

\begin{document}

\title{A single-photon router based on a modulated cavity optomechanical
system}
\author{Jun-Hao Liu}
\author{Ya-Fei Yu}
\email{yuyafei@m.scnu.edu.cn}
\author{Zhi-Ming Zhang}
\email{zhangzhiming@m.scnu.edu.cn}
\address{Guangdong Provincial Key Laboratory of Nanophotonic Functional Materials \& Devices (SIPSE), and
	Guangdong Provincial Key Laboratory of Quantum Engineering \& Quantum Materials, South China
	Normal University, Guangzhou 510006, China}
\begin{abstract}
We investigate the routing of a single-photon in a modulated cavity
optomechanical system, in which the cavity is driven by a strong coupling
field, and the mechanical resonator (MR) is modulated with a weak coherent
field. We show that, when there is no a weak coherent field modulating the
MR, the system cannot act as a single-photon router, since the signal will
be completely covered by the quantum and thermal noises. By introducing the
weak coherent field, we can achieve the routing of the single-photon by
adjusting the frequency of the weak coherent field, and the system can be
immune to the quantum and thermal noises.
\end{abstract}

\maketitle

\section{Introduction}

It is well known that 
quantum routers are important ingredients of quantum networks. 
In the past few decades, scientists have demonstrated that many physical
effects and physical systems, such as quantum interference \cite{1},
electromagnetically induced transparency \cite{2}, coupled waveguide array 
\cite{3,4,5}, can be used to realize the routing of photons. Recently, many
theoretical and experimental researches aiming at achieve the quantum router
in the single-photon level have been reported \cite{6,7,8,9,10}. Hall et al.
demonstrated the routing of single-photon without disturbing the photons'
quantum states with the help of strong cross-phase modulation \cite{11}. Hoi
et al. achieved a single-photon router in the microwave regime by using a
superconducting transmon qubit \cite{12}. Zhou et al. proposed an
experimentally accessible single-photon routing scheme using a three-level
atom embedded in a coupled-resonator waveguide \cite{13}.

We also notice that the realization of a single-photon router has been
researched in cavity optomechanical system. In Ref. \cite{14}, the authors
have shown how nanomechanical mirrors in an optical cavity can be used to
build single-photon routers. However, their analysis is inadequate. We find
that, their scheme actually cannot achieve the routing of a single-photon,
since the signal will be completely covered by the quantum and thermal
noises. In the present paper, we propose a scheme, based on a modulated
cavity optomechanical system, to realize the single-photon router. In our
system, the cavity is driven by a strong coupling field, and the mechanical
resonator (MR) is modulated with a weak coherent field. We can achieve the
routing of the single-photon by changing the frequency of the weak coherent
field. Moreover, our system can be immune to the quantum and thermal noises
when the MR is cooled to its quantum ground state.

\begin{figure}[t]
\centering\includegraphics[width=8cm,height=3.53cm]{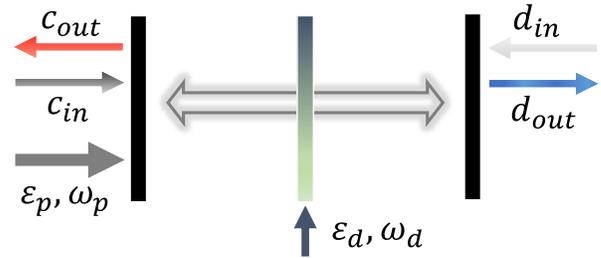}
\caption{(Color online) Schematic diagram of our proposed model. A
mechanical resonator (MR) of partial reflection is inserted between two
fixed mirrors. A strong coupling field is injected from the left. The MR is
modulated by a weak coherent field.}
\end{figure}

This paper is organized as follows. In Section II we introduce the
theoretical model. In Section III, we consider the case in which there is no
a weak coherent field modulating the MR. We show and explain why in this
case the system cannot act as a single-photon router. Next in Section IV, we
consider the case in which the MR is modulated by a weak coherent field. We
exhibit how the single-photon router works in this situation. We also
discuss the effects of the quantum and thermal noises on the single-photon
router. Finally in Section V, we provide a brief summary.

\section{Model}

Our proposed scheme is shown in Fig. 1. We consider a cavity optomechanical
system with a mechanical resonator (MR) of partial reflection inserted
between two fixed mirrors. The cavity is driven by a strong coupling field
with amplitude $\varepsilon _{p}=\sqrt{2\kappa P/(\hbar \omega _{p})}$ and
frequency $\omega _{p}$. The MR is modulated by a weak coherent field with
amplitude $\varepsilon _{d}$ and frequency $\omega _{d}$, this modulation
can be realized by, e.g., parametertically modulating the spring constant of
the MR at twice of the MR's resonance frequency \cite{15,16}. The
Hamiltonian of the system in the rotating frame at the frequency $\omega
_{p} $ of the coupling field is given by ($\hbar =1$) 
\begin{eqnarray}
H &=&\Delta \hat{a}^{\dag }\hat{a}+g_{0}\hat{a}^{\dag }\hat{a}(\hat{b}^{\dag
}+\hat{b})+i\varepsilon _{p}(\hat{a}^{\dag }-\hat{a})  \notag \\
&&+\omega _{m}\hat{b}^{\dag }\hat{b}+i\varepsilon _{d}[(\hat{b}^{\dag
})^{2}e^{-i2\omega _{d}t}-(\hat{b})^{2}e^{i2\omega _{d}t}],
\end{eqnarray}%
where $\Delta =\omega _{c}-\omega _{p}$ is the frequency detuning between
the cavity field and the coupling field. $\hat{a}$ and $\hat{b}$ are the
annihilation operators of the cavity mode and the mechanical mode with
frequency $\omega _{c}$ and $\omega _{m}$, respectively, $g_{0}$ is the
single-photon optomechanical coupling strength between the cavity mode and
the mechanical mode.

The system dynamics is fully described by the set of the quantum Langevin
equations (QLEs) 
\begin{eqnarray}
\frac{d\hat{a}}{dt} &=&-(2\kappa +i\Delta )\hat{a}-ig_{0}\hat{a}(\hat{b}%
^{\dag }+\hat{b})+\varepsilon _{p}  \notag \\
&&+\sqrt{2\kappa }\hat{c}_{in}+\sqrt{2\kappa }\hat{d}_{in}, \\
\frac{d\hat{b}}{dt} &=&-(\gamma +i\omega _{m})\hat{b}-ig_{0}\hat{a}^{\dag }%
\hat{a}+2\varepsilon _{d}e^{-i2\omega _{d}t}\hat{b}^{\dag }  \notag \\
&&+\sqrt{2\gamma }\hat{b}_{in},
\end{eqnarray}%
where $2\kappa $ is the total damping rate of the cavity and $\gamma $ is
the mechanical damping rate. $\hat{c}_{in}$, $\hat{d}_{in}$, and $\hat{b}%
_{in}$ are the input quantum fields with zero mean values.

We assume that the cavity field is driving by a strong coupling field $%
\varepsilon _{p}$ and the MR is modulated by a weak coherent field $%
\varepsilon _{d}$. The steady-state mean values of the operators can be
obtained from the QLEs (2)-(3) by making a transformations $\hat{a}%
\rightarrow \alpha +\delta \hat{a}$, and $\hat{b}\rightarrow \beta +\delta 
\hat{b}$, where $\alpha (\beta )$ and $\delta \hat{a}(\delta \hat{b})$ are
the steady state mean value and quantum fluctuation of the cavity mode
(mechanical mode), respectively, then we have 
\begin{eqnarray}
\alpha &=&\frac{\varepsilon _{p}}{2\kappa +i\Delta +ig_{0}(\beta +\beta
^{\ast })}, \\
\beta &=&\frac{-ig_{0}\left\vert \alpha \right\vert ^{2}}{\gamma +i\omega
_{m}}.
\end{eqnarray}

%
And for simplicity in symbols, we rewrite $\delta \hat{a}(\delta \hat{b})$
as $\hat{a}$($\hat{b}$) in the following sections.

\section{Without the weak coherent field}

In this section, we consider the case in which there is no a weak coherent
field modulating the MR. We would like to point out that this situation has
been discussed in Ref. \cite{14}. However, the analysis there is inadequate
and the conclusion is incorrect, so we make a re-calculation and a
re-discussion about this situation. When $\varepsilon _{d}$ $=$ $0$, the
linearized Hamiltonian of the system can be expressed as 
\begin{eqnarray}
H_{I} &=&\Delta ^{\prime }\hat{a}^{\dag }\hat{a}+\omega _{m}\hat{b}^{\dag }%
\hat{b}+G(\hat{a}^{\dag }\hat{b}+\hat{a}^{\dag }\hat{b}^{\dag })  \notag \\
&&+G^{\ast }(\hat{a}\hat{b}^{\dag }+\hat{a}\hat{b}),
\end{eqnarray}%
where $G$ $=$ $g_{0}\alpha $, $\Delta ^{\prime }$ $=$ $\Delta +g_{0}(\beta
+\beta ^{\ast })$ $\simeq $ $\Delta $.

We define a vector $v(t)$ $=$ ($\hat{a}(t)$, $\hat{b}(t)$, $\hat{a}^{\dag
}(t)$, $\hat{b}^{\dag }(t)$$)^{T}$ in terms of the operators of the system.
By substituting $v(t)$ and $H_{I}$ into the quantum Langevin equation, we
can obtain 
\begin{equation}
\frac{dv(t)}{dt}=Mv(t)+\sqrt{2\kappa }v_{c,in}+\sqrt{2\kappa }v_{d,in}+\sqrt{%
2\gamma }v_{b,in},
\end{equation}%
where $v_{x,in}$ $=$ $(\hat{x}_{in}(t)$, $0$, $\hat{x}_{in}^{\dag }(t)$, $%
0)^{T}$ ($x$ $=$ $c$, $d$), $v_{b,in}$ $=$ $(0$, $\hat{b}_{in}(t)$, $0$, $%
\hat{b}_{in}^{\dag }(t))^{T}$, and 
\begin{equation}
M=\left( 
\begin{array}{cccc}
-2\kappa -i\Delta  & -iG & 0 & -iG \\ 
-iG^{\ast } & -\gamma -i\omega _{m} & -iG & 0 \\ 
0 & iG^{\ast } & i\Delta -2\kappa  & iG^{\ast } \\ 
iG^{\ast } & 0 & iG & i\omega _{m}-\gamma 
\end{array}%
\right) .
\end{equation}%
The system is stable only when the real parts of all the eigenvalues of
matrix $M$ are negative. The stability conditions can be explicitly given by
using the Routh-Hurwitz criterion \cite{17,18,19}, and the stability
conditions are fulfilled in the system with our used parameters. Moreover,
for simplicity, we take $G$ as a real number in the following calculations.

\begin{figure*}[tbp]
\centering\includegraphics[width=17cm,height=8.76cm]{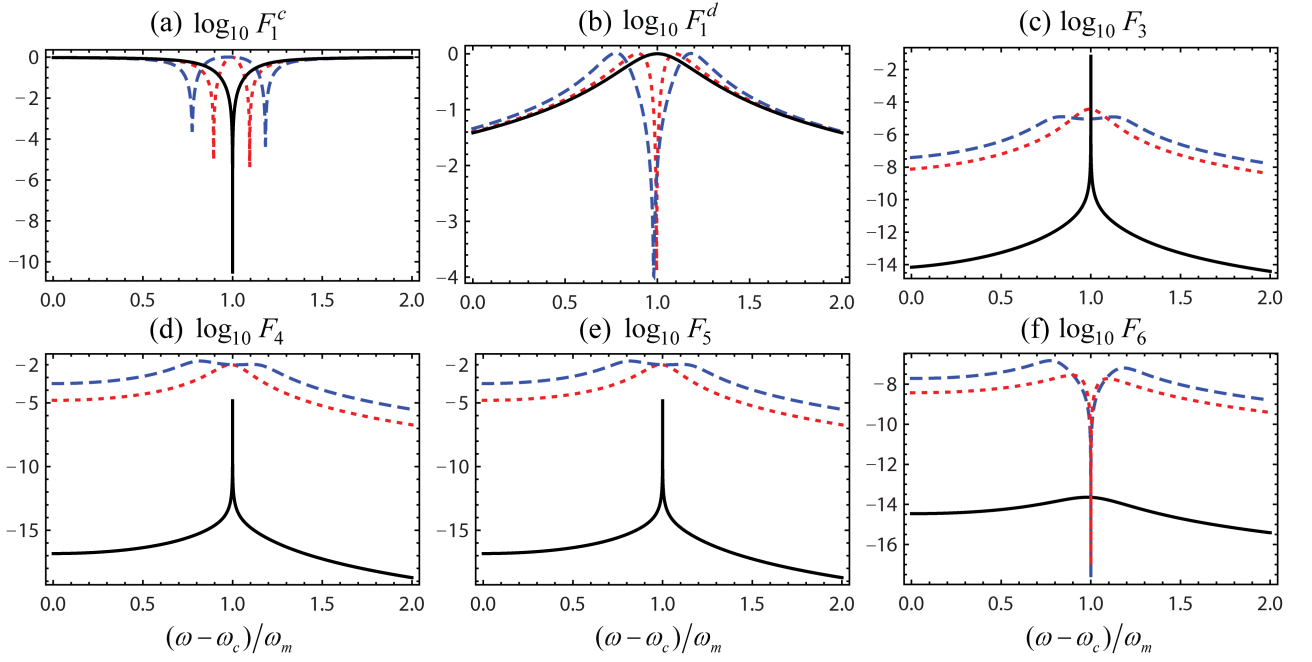}
\caption{(Color online) The spectrums of the scattering probabilities $%
F_{1}^{c}$, $F_{1}^{d}$, $F_{3}$, $\cdots $, $F_{6}$ for $G/\protect\omega %
_{m}$ $=$ $10^{-4}$ (black solid lines), $G/\protect\omega _{m}$ $=$ $0.1$
(red dotted lines), $G/\protect\omega _{m}$ $=$ $0.2$ (blue dashed lines).
The other parameters are stated in the text.}
\end{figure*}

In experiments, the fluctuations of the electromagnetic field are more
convenient to be measured in the frequency domain than in the time domain.
Therefore, we introduce the Fourier transform of the operators 
\begin{eqnarray}
\hat{o}(\omega ) &=&\int_{-\infty }^{+\infty }\hat{o}(t)e^{i\omega t}dt, \\
\hat{o}^{\dag }(\omega ) &=&\int_{-\infty }^{+\infty }\hat{o}^{\dag
}(t)e^{i\omega t}dt,
\end{eqnarray}%
where $\hat{o}=\hat{a}$, $\hat{b}$, then we can solve the linearized QLEs
(7) in the frequency domain 
\begin{eqnarray}
v(\omega ) &=&-(M+i\omega I)^{-1}[\sqrt{2\kappa }v_{c,in}(\omega )  \notag \\
&&+\sqrt{2\kappa }v_{d,in}(\omega )+\sqrt{2\gamma }v_{b,in}(\omega )],
\end{eqnarray}%
where $v(\omega )$ $=$ $($$\hat{a}(\omega )$, $\hat{b}(\omega )$, $\hat{a}%
^{\dag }(\omega )$, $\hat{b}^{\dag }(\omega ))^{T}$, $v_{x,in}(\omega )$ $=$ 
$(\hat{x}_{in}(\omega )$, $0$, $\hat{x}_{in}^{\dag }(\omega )$, $0)^{T}$ ($x$
$=$ $c,d$), and $v_{b,in}(\omega )$ $=$ $($$0$, $\hat{b}_{in}(\omega )$, $0$%
, $\hat{b}_{in}^{\dag }(\omega ))^{T}$, then we can obtain 
\begin{equation}
\hat{a}(\omega )=\mathit{f}(\omega )v_{in}(\omega ),
\end{equation}%
where $\mathit{f}(\omega )$ $=$ $($$f_{1}(\omega )$, $f_{2}(\omega )$, $%
f_{3}(\omega )$, $f_{4}(\omega )$, $f_{5}(\omega )$, $f_{6}(\omega ))$, and $%
v_{in}(\omega )$ $=$ $(\hat{c}_{in}(\omega )$, $\hat{d}_{in}(\omega )$, $%
\hat{b}_{in}(\omega )$, $\hat{c}_{in}^{\dag }(\omega )$, $\hat{d}_{in}^{\dag
}(\omega )$, $\hat{b}_{in}^{\dag }(\omega ))^{T}$, the concrete forms of the
coefficients $f_{1}(\omega )$, $\cdots $, $f_{6}(\omega )$ are tediously
long, we will not write them out here.

In this paper, we consider that the input field $\hat{c}_{in}$ is in a
single-photon Fock state, and the correlation functions are $\left\langle 
\hat{c}_{in}^{\dag }(\Omega )\hat{c}_{in}(\omega )\right\rangle $ $=$ $%
S_{in}(\omega )\delta (\omega +\Omega )$, $\left\langle \hat{c}_{in}(\Omega )%
\hat{c}_{in}^{\dag }(\omega )\right\rangle $ $=$ $[S_{in}(\Omega )+1]\delta
(\omega +\Omega )$. It should be point out that, when we use such a
single-photon state as the input state to the cavity, we also assume that
its center frequency is resonant with the cavity \cite{20,21}. Its spectrum
is given by the Lorentzian lineshape $S_{in}(\omega )$ $=$ $\frac{\Gamma
/\pi }{(\omega -\omega _{c})^{2}+\Gamma ^{2}}$, in which $\Gamma $ is the
decay rate of the single photon. The incoming vacuum field $\hat{d}_{in}$ is
characterized by $\left\langle \hat{d}_{in}(\Omega )\hat{d}_{in}^{\dag
}(\omega )\right\rangle $ $=$ $\delta (\Omega +\omega )$. The mechanical
input operator $\hat{b}_{in}$ satisfies $\left\langle \hat{b}_{in}^{\dag
}(\Omega )\hat{b}_{in}(\omega )\right\rangle $ $=$ $n_{th}\delta (\Omega
+\omega )$, $\left\langle \hat{b}_{in}(\Omega )\hat{b}_{in}^{\dag }(\omega
)\right\rangle $ $=$ $(n_{th}+1)\delta (\Omega +\omega )$ in the frequency
domain, where $n_{th}$ is the thermal phonon occupation number at a finite
temperature

The relation among the input, internal, and output fields is given as \cite%
{22} 
\begin{equation}
\hat{x}_{out}(\omega )=-\hat{x}_{in}(\omega )+\sqrt{2\kappa }\hat{a}(\omega
),x=c,d.
\end{equation}%
Then we can write the operators of the output fields as 
\begin{equation}
\hat{c}_{out}(\omega )=\mathit{f}^{c}(\omega )v_{in}(\omega ),\hat{d}%
_{out}(\omega )=\mathit{f}^{d}(\omega )v_{in}(\omega ),
\end{equation}%
where $\mathit{f}^{c}(\omega )$ $=$ $(f_{1}^{\prime }(\omega )$ $-$ $1$, $%
f_{2}^{\prime }(\omega )$, $f_{3}^{\prime }(\omega )$, $f_{4}^{\prime
}(\omega )$, $f_{5}^{\prime }(\omega )$, $f_{6}^{\prime }(\omega ))$, $%
\mathit{f}^{d}(\omega )$ $=$ $(f_{1}^{\prime }(\omega )$, $f_{2}^{\prime
}(\omega )-1$, $f_{3}^{\prime }(\omega )$, $f_{4}^{\prime }(\omega )$, $%
f_{5}^{\prime }(\omega )$, $f_{6}^{\prime }(\omega ))$, and $f_{j}^{\prime
}(\omega )$ $\equiv $ $\sqrt{2\kappa }f_{j}(\omega )$ ($j=1,2,3,4,5,6$).

The spectrums of the output fields are defined by 
\begin{equation}
S_{x,out}(\omega )=\int d\Omega \left\langle \hat{x}_{out}^{\dag }(\Omega )%
\hat{x}_{out}(\omega )\right\rangle ,x=c,d.
\end{equation}%
By substituting the expressions of $\hat{c}_{out}(\omega )$ and $\hat{d}%
_{out}(\omega )$ into Eq. (15), and using the correlation functions, one can
obtain 
\begin{eqnarray}
S_{c,out}^{I}(\omega ) &=&F_{1}^{c}S_{in}(\omega )+F_{3}n_{th}  \notag \\
&&+F_{4}S_{vac}(-\omega )+F_{5}+F_{6}, \\
S_{d,out}^{I}(\omega ) &=&F_{1}^{d}S_{in}(\omega )+F_{3}n_{th}  \notag \\
&&+F_{4}S_{vac}(-\omega )+F_{5}+F_{6},
\end{eqnarray}%
where $F_{1}^{c}$ $=\left\vert f_{1}^{\prime }(\omega )-1\right\vert ^{2}$, $%
F_{1}^{d}$ $=\left\vert f_{1}^{\prime }(\omega )\right\vert ^{2}$, $F_{3}$ $%
= $ $\left\vert f_{3}^{\prime }(\omega )\right\vert ^{2}$ $+$ $\left\vert
f_{6}^{\prime }(\omega )\right\vert ^{2}$, $F_{j}$ $=$ $\left\vert
f_{j}^{\prime }(\omega )\right\vert ^{2}$ ($j$ $=$ $4,5,6$), and $%
S_{vac}(-\omega )$ $=$ $S_{in}(-\omega )+1$.

\begin{figure*}[tbp]
\centering\includegraphics[width=17cm,height=8.76cm]{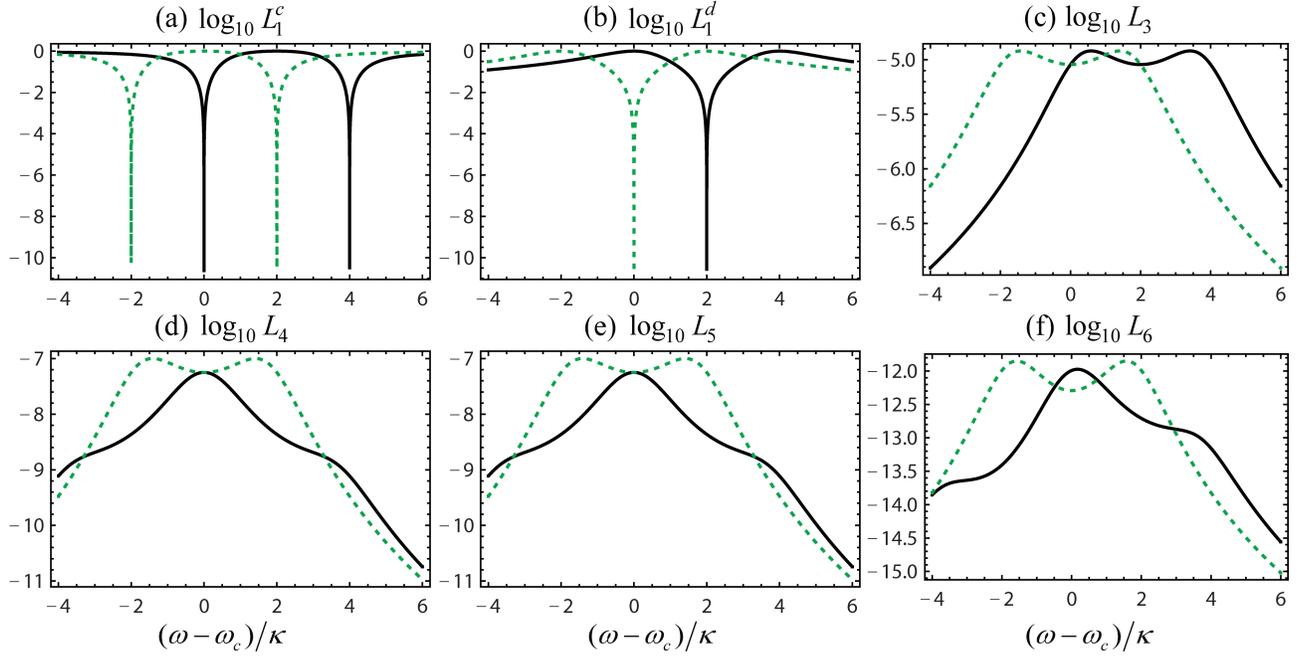}
\caption{(Color online) The spectrums of the scattering probabilities $%
L_{1}^{c}$, $L_{1}^{d}$, $L_{3}$, $\cdots $, $L_{6}$ for $\protect\omega %
_{d}/\protect\omega _{m}$ $=$ $0.8$ (black solid lines), $\protect\omega %
_{d}/\protect\omega _{m}$ $=$ $1$ (green dotted lines). The amplitude of the
weak coherent field $\protect\varepsilon _{d}/\protect\kappa $ $=$ $%
2.37\times 10^{-4}$, and the other parameters are stated in the text.}
\end{figure*}

We can see that both the spectrums $S_{c,out}^{I}(\omega )$ and $%
S_{d,out}^{I}(\omega )$ of the output fields contain five components. For $%
S_{c,out}^{I}(\omega )$, $F_{1}^{c}$ and $F_{4}$ represent, respectively,
the scattering probabilities of the input field $\hat{c}_{in}(\omega )$ and
its fluctuation. $F_{5}$ is the scattering probability of the fluctuation of
the vacuum field $\hat{d}_{in}(\omega )$. $F_{3}$ and $F_{6}$ denote,
respectively, the scattering probabilities of the thermal noise and quantum
noise input to the mechanical mode. It can be seen that even if there is no
input signal photon, the output fields will also be generated by the vacuum
fluctuations and thermal noises. A good single photon router should not be
influenced by these quantum and thermal noises.

Next we numerically calculate the reflection spectrum $S_{c,out}^{I}(\omega
) $ and the transmission spectrum $S_{d,out}^{I}(\omega )$. The parameters
we used are the same as that in Ref. \cite{14}: $\lambda $ $=$ $1054$ nm, $L$
$= $ $6.7$ cm, $m$ $=$ $40$ ng, $\omega _{m}$ $=$ $2\pi \times 134$ kHz, $%
\gamma _{m}$ $=$ $0.76$ Hz, $\kappa $ $=$ $0.1\omega _{m}$, $\Delta $ $=$ $%
\omega _{m}$.

First we do not consider the effects of the noises. In Fig. 2(a) and Fig.
2(b), we plot the spectrums of the scattering probabilities $F_{1}^{c}$ and $%
F_{1}^{d}$ for different effective optomechanical coupling strength $G$. It
can be seen that, when $G$ is small ($G$ $=$ $10^{-4}\omega _{m}$) \cite{23}%
, around the point $\omega -\omega _{c}$ $=$ $\omega _{m}$, the spectrum $%
F_{1}^{c}$ exhibits a valley, while the spectrum $F_{1}^{d}$ exhibits a
peak, and at $\omega -\omega _{c}$ $=$ $\omega _{m}$, we have $F_{1}^{c}$ $%
\approx $ $0$ and $F_{1}^{d}$ $\approx $ $1$, this means that the
single-photon will completely transmit through the cavity and exit from the
right output port. For a larger $G$, e.g., for $G$ $=$ $0.1\omega _{m}$ or $%
0.2\omega _{m}$, at $\omega -\omega _{c}$ $=$ $\omega _{m}$, we have $%
F_{1}^{c}$ $\approx $ $1$ and $F_{1}^{d}$ $\approx $ $0$, this means that
the single-photon will be completely reflected from the cavity and exit from
the left output port. These results are similar with that in Ref. \cite{14},
and indicate that one can realize a single-photon router by adjusting the
effective optomechanical coupling strength $G$. However, it should be
pointed out that one obtains these results when one ignores the noises.
Things will be different if the noises are taken into accounted, and this
will be discussed in the following. Here we would like to point out another
phenomenon: the spectrums will exhibit a split at $\omega -\omega _{c}$ $=$ $%
\omega _{m}$ $\pm $ $G$, and this is associated with the normal mode
splitting \cite{24,25}.

Now we estimate the order of magnitude of the signal. It can be seen that,
in this case, the operating frequency of the system is at $\omega -\omega
_{c}$ $=$ $\omega _{m}$, hence the signal can be expressed as $S_{in}(\omega
_{c}+\omega _{m})$ $=$ $\frac{\Gamma /\pi }{(\omega _{m})^{2}+\Gamma ^{2}}$.
Its maximum value is about $10^{-7},$ which appears at $\Gamma $ $=$ $\omega
_{m}$.

Then we consider the effects of the quantum and thermal noises. In Fig.
2(c)-(f), we plot the spectrums of the scattering probabilities $F_{3}$, $%
\cdots $, $F_{6}$, respectively. We find that, at $\omega -\omega _{c}$ $=$ $%
\omega _{m}$, $F_{3}$, $\cdots $, $F_{6}$ have the following order of
magnitudes: when $G$ $=$ $10^{-4}\omega _{m}$, one has $F_{3}$ $\sim $ $10^{-2}$, $%
F_{4}$ $\sim $ $F_{5}$ $\sim $ $10^{-5}$, and $F_{6}$ $\sim $ $10^{-14}$, when $%
G $ $=$ $0.1\omega _{m}$ or $0.2\omega _{m}$, one has $F_{3}$ $\sim $ $10^{-5}$, $%
F_{4} $ $\sim $ $F_{5}$ $\sim $ $10^{-2}$, and $F_{6}$ $\sim $ $10^{-17}$. That
is, with the increase of $G$, the noises deriving from the input fields $%
\hat{c}_{in}(\omega )$ and $\hat{d}_{in}(\omega )$ have been strongly
amplified, while the noises deriving from the input field $\hat{b}%
_{in}(\omega )$ has been effectively suppressed.

By comparing the order of magnitudes of the signal and the noises, we can
see that, in the reflection spectrum $S_{c,out}^{I}(\omega )$ and the
transmission spectrum $S_{d,out}^{I}(\omega )$, the contributions $%
F_{1}^{c}S_{in}(\omega _{c}+\omega _{m})$ or $F_{1}^{d}S_{in}(\omega
_{c}+\omega _{m})$ from the signal is much less than the contributions $%
F_{3}n_{th}$ $+$ $F_{4}$ $+$ $F_{5}$ from the noises, whether $G$ $=$ $%
10^{-4}\omega _{m}$ , $0.1\omega _{m}$, or $0.2\omega _{m}$. In other words,
the signal will be completely covered by the quantum and thermal noises.
Hence, we can conclude that, in this case, this system can not act as a
single-photon router.

\section{With the weak coherent field}

In this section, we consider the case in which there is a weak coherent
field modulating the MR. In the rotation frame with $H^{\prime }$ $=$ $%
\omega _{d}(\hat{a}^{\dag }\hat{a}+\hat{b}^{\dag }\hat{b})$, the linearized
Hamiltonian of the system can be expressed as 
\begin{eqnarray}
H_{II} &=&\delta \hat{a}^{\dag }\hat{a}+\Delta _{m}\hat{b}^{\dag }\hat{b}+G%
\hat{a}^{\dag }\hat{b}+G^{\ast }\hat{a}\hat{b}^{\dag }  \notag \\
&&+i\varepsilon _{d}[(\hat{b}^{\dag })^{2}-(\hat{b})^{2}],
\end{eqnarray}%
where $\delta =\Delta -\omega _{d}$, and $\Delta _{m}=\omega _{m}-\omega
_{d} $. Here we have used the rotating-wave approximation to omit the
high-frequency oscillation terms $\hat{a}^{\dag }\hat{b}^{\dag }e^{i2\omega
_{d}t}$ and $\hat{a}\hat{b}e^{-i2\omega _{d}t}$.

By substituting $v(t)$ and $H_{II}$ into the quantum Langevin equation, we
can obtain 
\begin{equation}
\frac{dv(t)}{dt}=M^{\prime }v(t)+\sqrt{2\kappa }v_{c,in}+\sqrt{2\kappa }%
v_{d,in}+\sqrt{2\gamma }v_{b,in},
\end{equation}%
where 
\begin{equation}
M^{\prime }=\left( 
\begin{array}{cccc}
-2\kappa -i\delta & -iG & 0 & 0 \\ 
-iG^{\ast } & -\gamma -i\Delta _{m} & 0 & 2\varepsilon _{d} \\ 
0 & 0 & i\delta -2\kappa & iG^{\ast } \\ 
0 & 2\varepsilon _{d} & iG & i\Delta _{m}-\gamma%
\end{array}%
\right) .
\end{equation}%
The stability conditions of the matrix $M^{\prime }$ have been verified by
using the Routh-Hurwitz criterion with our used parameters. The subsequent
calculations are similar with that in section III. The spectrums of the
output fields are obtained as 
\begin{eqnarray}
S_{c,out}^{II}(\omega ) &=&L_{1}^{c}S_{in}(\omega )+L_{3}n_{th}  \notag \\
&&+L_{4}S_{vac}(-\omega )+L_{5}+L_{6}, \\
S_{d,out}^{II}(\omega ) &=&L_{1}^{d}S_{in}(\omega )+L_{3}n_{th}  \notag \\
&&+L_{4}S_{vac}(-\omega )+L_{5}+L_{6},
\end{eqnarray}%
in which $L_{1}^{c}$, $L_{1}^{d}$, $L_{3}$, $\cdots $, $L_{6}$ have the same
physical meaning with $F_{1}^{c}$, $F_{1}^{d}$, $F_{3}$, $\cdots $, $F_{6}$,
respectively. The concrete forms of $L_{1}^{c}$, $L_{1}^{d}$, $L_{3}$, $%
\cdots $, $L_{6}$ are too verbose to be given here.

\begin{figure}[b]
\centering\includegraphics[width=8cm,height=6cm]{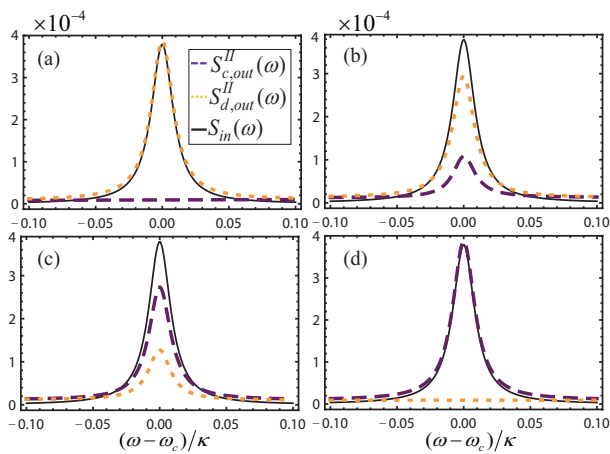}
\caption{(Color online) The spectrums of the output fields $S_{c,out}^{II}(%
\protect\omega )$ (purple dashed lines), $S_{d,out}^{II}(\protect\omega )$
(orange dotted lines) and the input fields $S_{in}(\protect\omega )$ (black
solid lines) for different $\protect\omega _{d}$: (a) $\protect\omega _{d}/%
\protect\omega _{m}$ $=$ $0.8$, (b) $\protect\omega _{d}/\protect\omega _{m}$
$=$ $0.85$, (c) $\protect\omega _{d}/\protect\omega _{m}$ $=$ $0.9$, (d) $%
\protect\omega _{d}/\protect\omega _{m}$ $=$ $1$. The thermal phonon
occupation number $n_{th}$ $=$ $1$, and the other parameters are the same as
in Fig. 3.}
\end{figure}

Let us show how the single-photon router works in our scheme. In our system,
with the existence of the weak coherent field, the effective frequency of
the mechanical mode becomes $\Delta _{m}=\omega _{m}-\omega _{d}$. Figures
3(a) and 3(b) show the spectrums of the scattering probabilities $L_{1}^{c}$
and $L_{1}^{d}$, in which we have chosen $G$ $=$ $0.2\omega _{m}$ $=$ $%
2\kappa $. It can be seen that, when $\omega _{d}$ $=$ $0.8\omega _{m}$ ($%
\Delta _{m}=0.2\omega _{m}=2\kappa $), $L_{1}^{c}$ exhibit a peak around the
point $\omega -\omega _{c}$ $=$ $\Delta _{m}=2\kappa $ and two valleys
around the points $\omega -\omega _{c}$ $=$ $\Delta _{m}\pm G=\left(
0,4\kappa \right) $, while $L_{1}^{d}$ exhibits just the opposite. At $%
\omega -\omega _{c}$ $=$ $0$, one has $L_{1}^{c}$ $\approx $ $0$ and $%
L_{1}^{d}$ $\approx $ $1$, this means that the single-photon will completely
transmit through the cavity and exit from the right output port. With the
increase of $\omega _{d}$, the curves will integrally move to the left, for
example, when $\omega _{d}$ $=$ $\omega _{m}$ ($\Delta _{m}=0$), we have $%
L_{1}^{c}$ $\approx $ $1$ and $L_{1}^{d}$ $\approx $ $0$ at $\omega -\omega
_{c}$ $=$ $0$, this means that the single-photon will be completely
reflected from the cavity and exit from the left output port. In this way,
we can achieve the routing of the single-photon. If we consider that the
single-photon has a Lorentzian lineshape with a narrower linewidth than the
cavity ($\Gamma $ $=$ $0.01\kappa $), we can estimate that the signal $%
S_{in}(\omega _{c})$ has the order of magnitude $10^{-4}$.

Now we consider the effects of the quantum and thermal noises in this case.
In Fig. 3(c)-(f), we plot the spectrums of the scattering probabilities $%
L_{3}$, $\cdots $, $L_{6}$, respectively. We find that, in the range of the
parameters we considered ($\varepsilon _{d}/\kappa $ $=$ $2.37\times
10^{-4}$), at $\omega -\omega _{c}$ $=$ $0$, $L_{3}$, $\cdots $, $L_{6}$
have the following order of magnitudes: when $\omega _{d}$ $=$ $0.8\omega
_{m}$ or $\omega _{m}$, one has $L_{3}$ $\sim $ $10^{-5}$, $L_{4}\sim L_{5}$ 
$\sim $ $10^{-8}$, and $L_{6}$ $\sim $ $10^{-12}$.

By comparing with the case in which there is no a weak coherent field, we
find that, in the present case, the signal is enhanced, and the noises are
suppressed. By comparing the order of magnitudes of the signal and the
noises, we can write the spectrum of the output fields as%
\begin{eqnarray}
S_{c,out}^{II}(\omega ) &\approx &L_{1}^{c}S_{in}(\omega )+L_{3}n_{th}, \\
S_{d,out}^{II}(\omega ) &\approx &L_{1}^{d}S_{in}(\omega )+L_{3}n_{th}.
\end{eqnarray}%
It can be seen that if the thermal phonon occupation number $n_{th}$ $%
\lesssim $ $1$, the signal can not be covered by the noises.

In Fig. 4 we plot the spectrums of the output fields $S_{d,out}^{II}(\omega
) $ and $S_{c,out}^{II}(\omega )$ for different $\omega _{d}$. For $\omega
_{d} $ $=$ $0.8\omega _{m}$ and at $\omega -\omega _{c}$ $=$ $0,$ we find $%
S_{d,out}^{II}(\omega _{c})$ $\approx $ $S_{in}(\omega _{c})$, and $%
S_{c,out}^{II}(\omega _{c})$ $\approx $ $0$. If we increase $\omega _{d}$, $%
S_{d,out}^{II}(\omega )$ will gradually decrease, and $S_{c,out}^{II}(\omega)
$ will gradually increase near $\omega -\omega _{c}$ $=$ $0$. When $\omega
_{d}$ $=$ $\omega _{m}$, we have $S_{c,out}^{II}(\omega _{c})$ $\approx $ $%
S_{in}(\omega _{c})$, $S_{d,out}^{II}(\omega _{c})$ $\approx $ $0$. This
shows that our system can act as a single-photon router.

\section{Conclusion}

In summary, we have investigated the routing of a single-photon in a
modulated cavity optomechanical system, in which the cavity is driven by a
strong coupling field, and the mechanical resonator is modulated by a weak
coherent field. We have shown that, if there is no the weak coherent field,
the signal will be completely covered by the quantum and thermal noises, and
the single-photon router cannot be realized. By introducing a weak coherent
field modulating the mechanical resonator, we can achieve the single-photon
router by adjusting the frequency of the weak coherent field.

\section*{ACKNOWLEDGEMENTS}

\addcontentsline{toc}{section}{Acknowledgements} This work was supported by
the National Natural Science Foundation of China (Nos. 11574092, 61775062,
61378012, 91121023); the National Basic Research Program of China (No.
2013CB921804); the Innovation Project of Graduate School of South China
Normal University.


\begin{thebibliography}{99}
\bibitem{23} It should be point out that $G$ cannot be zero, since $G$ $=$ $%
g_{0}\alpha $, by definition $g_{0}$ $\neq $ $0$, and the linearization of
the Hamiltonian (from Eq. (1) to Eq. (6)) demands $\alpha $ $\neq $ $0$ (and
in turn $\varepsilon _{p}$ $\neq $ $0$, $P$ $\neq $ $0$ ). In fact, the
linearization demands $\left\vert \alpha \right\vert $ $\gg $ $1$, i.e. $G$ $%
\gg $ $g_{0}$. For the parameters here, $g_{0}$ is in the order of $10$ Hz,
so we take $G$ $=$ $10^{-4}\omega _{m}$ $\approx $ $10^{2}$ Hz.
\end{thebibliography}
\end{document}